\documentstyle[psfig]{l-aa}		
\def\CO10{{\hbox {CO(1--0)}}}

\def\,{\thinspace}
\def\Msun{M$_\odot$}

\def\Lsun{L$_\odot$}

\def \kms{km\,s$^{-1}$}

\begin{document}
\thesaurus{11 (11.09.1 HDF 850.1, 11.09.4, 11.11.1, 11.19.6)}
\title{
Proposed identification of Hubble Deep Field submillimeter source HDF~850.1
}
\author{
D.~Downes\inst{1} 
\and R.~Neri\inst{1}
\and A.~Greve\inst{1}  
\and S.~Guilloteau\inst{1} 
\and F.~Casoli\inst{2}
\and D.~Hughes\inst{3,4}
\and D.~Lutz\inst{5}
\and K.M.~Menten\inst{6}
\and D.J.~Wilner\inst{7}
\and P.~Andreani\inst{8}
\and F.~Bertoldi\inst{6}
\and C.L.~Carilli\inst{9}
\and J.~Dunlop\inst{3}
\and R.~Genzel\inst{5}
\and F.~Gueth\inst{6}
\and R.J.~Ivison\inst{10}
\and R.G. Mann\inst{11}
\and Y.~Mellier\inst{2,12}
\and S.~Oliver\inst{11}
\and J.~Peacock\inst{3}
\and D.~Rigopoulou\inst{5}
\and M.~Rowan-Robinson\inst{11}
\and P.~Schilke\inst{6}
\and S.~Serjeant\inst{11}
\and L.J.~Tacconi\inst{5}
\and M.~Wright\inst{13}
}
\institute{Institut de Radio Astronomie Millim\'etrique, Domaine Universitaire,
           F-38406 St. Martin d'H\`eres,  France
\and DEMIRM, Observatoire de Paris, 61 av. de l'Observatoire, 
	F-75014 Paris, France, and UMR 8540 du CNRS
\and Institute for Astronomy, University of Edinburgh, Royal Observatory, 
	Blackford Hill, Edinburgh, EH9 3HJ, UK
\and Instituto Nacional de Astrofisica, Optica y Electronica (INAOE),
	Apartado Postal 51 y 216, 72000 Puebla, Pue., Mexico
\and Max-Planck-Institut f\"ur extraterrestrische Physik, D-85748
	Garching-bei-M\"unchen, Germany
\and Max-Planck-Institut f\"ur Radioastronomie, Auf dem H\"ugel 69, 
	D-53121 Bonn, Germany
\and Center for Astrophysics, 60 Garden St., Cambridge, MA 02138, USA
\and Dipartimento di Astronomia, Universit\`a di Padova, 
vicolo dell'Osservatorio 5, I-35122 Padova, Italy
\and National Radio Astronomy Observatory, P.O. Box O, Socorro, N.M., 87801, USA
\and Dept. of Physics and Astronomy, University College London, Gower Street,
	London, WC1E 6BT, UK
\and Astrophysics Group, Imperial College London, Blackett Laboratory,
	Prince Consort Road, London SW7 2BZ, UK
\and Institut d'Astrophysique, 98bis, Bd Arago, 75014, Paris, France
\and Radio Astronomy Laboratory, University of California, Berkeley, 
CA94720, USA
}

\offprints{D.~Downes}
\date{received date; accepted date\\}
\maketitle
\begin{abstract}
The IRAM Interferometer has been used to detect the  submillimeter source HDF
850.1 found by Hughes et al. (1998) in the Hubble Deep Field. The flux density
measured at 1.3\,mm (236\,GHz) is 2.2$\pm 0.3 (1\sigma)$\,mJy,  in agreement
with the flux  density measured  at the JCMT.    The flux densities and upper
limits measured at 3.4, 2.8, 1.3, 0.85,  and 0.45\,mm show that the emission
comes from dust.
We suggest that the 1.3\,mm dust source is associated with the optical 
 arc-like feature, 3-593.0, that
has a photometric redshift $z\approx 1.7$.  
If the HDF~850.1 is at this redshift and unlensed, its spectral energy 
distribution, combined with that of 3-593.0, matches closely that
of the ultraluminous galaxy VII~Zw~31.  Another possibility is that 
the dust source may be gravitationally lensed 
by the  elliptical galaxy 3-586.0 at $z\approx 1$.    

The position of the dust source agrees 
within the errors with that of the tentative 
VLA radio source 3651+1226.
\end{abstract}

\section{Introduction}
Hughes et al. (1998) observed the Hubble Deep Field with the SCUBA multi-beam
bolometer array at the James Clerk Maxwell Telescope (JCMT) at 850 and
450\,$\mu$m.  Apart from  HDF 850.4 (and possibly HDF 850.3) the SCUBA survey
did not detect  the  mid-IR sources found in the Hubble Deep Field with the ISO
satellite  (Rowan-Robinson et al. 1997; Mann et al. 1997; Aussel et al. 1999).
In general these ISOCAM detections are associated with $z<1$ galaxies, while the
submillimeter survey appears to  have revealed a higher-redshift population of
dust enshrouded starburst galaxies. Further SCUBA surveys of other fields 
(e.g., 
Barger et al. 1999a,b; Eales et al. 1999; Lilly et al. 1999) indicate 
however that many of the sub-mm sources have redshifts smaller than 1, 
and that much of the star-forming activity in galaxies 
has occurred relatively 
recently, at $z\sim 2$.

\begin{table*}
\caption{Source Positions}
\begin{tabular}{l c c c c c c}
\hline
 		&R.A. 	  &Dec.		&$\Delta \theta$ from  	&Redshift	
					&Remark			&Refs.
\\ 
{\bf Source} 	&(J2000)  &(J2000)	&HDF850.1	&$z$ & 
\\ 			     			     
\hline
\\
\multicolumn{5}{l}{ {\bf Proposed identification:} }
\\
HDF-850.1   
	&12$^{\rm h}$36$^{\rm m}$51.98$^{\rm s} \pm 0.04^{\rm s}$   	
	&62$^\circ12'25.7'' \pm 0.3''$ 		
	&---
	&---
	&1.3\,mm dust source 		
	&1  	
	\\
3-593.0
	&12$^{\rm h}$36$^{\rm m}$51.86$^{\rm s} \pm 0.06^{\rm s}$
	&62$^\circ12'25.6'' \pm 0.4''$ 
	&$0.8''$
	&1.73 to 1.76
	&``arc 1" 
	&2, 6, 7
\\
3-593.2
	&12$^{\rm h}$36$^{\rm m}$51.94$^{\rm s} \pm 0.06^{\rm s}$
	&62$^\circ12'25.0'' \pm 0.4''$ 
	&$0.8''$
	&1.82
	&``arc 2" 
	&2, 6
\\
3651+1226
        &12$^{\rm h}$36$^{\rm m}$51.96$^{\rm s} \pm 0.14^{\rm s}$ 
 	&62$^\circ12'26.1'' \pm 1.0''$		
	&$0.4''$
	&---
	&VLA (3\,cm) ?		
	&3
\\
\\
\multicolumn{5}{l}{ {\bf Other nearby optical sources within 3$''$:} }
\\
3-586.0
	&12$^{\rm h}$36$^{\rm m}$52.10$^{\rm s} \pm 0.06^{\rm s}$
	&62$^\circ 12'26.3'' \pm 0.4''$ 
	&$1.0''$
	&1.0 to 1.2
	&elliptical 
	&2, 6, 7
\\
3-577.0
	&12$^{\rm h}$36$^{\rm m}$52.26$^{\rm s} \pm 0.06^{\rm s}$
	&62$^\circ 12'27.2'' \pm 0.4''$ 
	&$2.5''$
	&2.88, 2.89, 3.36 ?
	&``counterimage" 
	&6, 7, 8
\\
3-613.0	&12$^{\rm h}$36$^{\rm m}$52.37$^{\rm s} \pm 0.06^{\rm s}$
	&62$^\circ 12'25.1'' \pm 0.4''$
	&$2.8''$
	&1.64
	&blue irregular 		
	&2, 6, 7
\\
\\
\multicolumn{5}{l}{ {\bf Other nearby sources within 6$''$:} }
\\
3-550.2	&12$^{\rm h}$36$^{\rm m}$51.34$^{\rm s} \pm 0.06^{\rm s}$
	&62$^\circ 12'26.9'' \pm 0.4''$
	&$4.6''$
	&1.72
	&blue irregular 
	&2, 6, 7
\\
3-633.1
	&12$^{\rm h}$36$^{\rm m}$51.60$^{\rm s} \pm 0.06^{\rm s}$
	&62$^\circ 12'22.5'' \pm 0.4''$ 
	&$4.2''$
	&1.72
	&irregular ?
	&2, 6, 7
\\
3-659.1
	&12$^{\rm h}$36$^{\rm m}$51.72$^{\rm s} \pm 0.06^{\rm s}$
	&62$^\circ 12'20.2'' \pm 0.4''$ 
	&$5.8''$
	&0.299
	&Sb spiral (Cowie 1999)
	&2, 6
\\
3651+1221
        &12$^{\rm h}$36$^{\rm m}$51.65$^{\rm s} \pm 0.02^{\rm s}$ 
 	&62$^\circ12'21.4'' \pm 0.2''$		
	&$4.9''$
	&---
	&VLA (20\,cm, 3\,cm)		
	&3
\\
PM3\_29
&12$^{\rm h}$36$^{\rm m}$51.9$^{\rm s}\phantom{0} \pm 0.4^{\rm s}\phantom{0}$ 
 	&62$^\circ12'21''\phantom{.0} \pm 3''\phantom{.0}$		
	&$4.7''$
	&---
	&ISOCAM (15\,$\mu$m)		
	&4
\\
tentative?
        &12$^{\rm h}$36$^{\rm m}$51.58$^{\rm s} \pm 0.06^{\rm s}$ 
 	&62$^\circ12'20.5'' \pm 0.4''$		
	&$5.9''$
	&---
	&3$\sigma$ contour at 1.3\,mm		
	&1
\\
\\
\multicolumn{3}{l}{ {\bf Adopted positions for phase calibrators:} }
	&&{\bf Flux:} &{\bf 1.3\,mm}\ \ \ \ \ \ {\bf 3.4\,mm}
\\
1044+719
        &10$^{\rm h}$48$^{\rm m}$27.620$^{\rm s} \pm 0.002^{\rm s}$ 
 	&71$^\circ 43'35.93'' \pm 0.01''$		
	&$14.0^\circ$
	&---
	&0.39\,Jy\ \ \ \ \ \ \ \ \ 0.68\,Jy 
	&5		
\\
1125+596
	&11$^{\rm h}$28$^{\rm m}$13.342$^{\rm s} \pm 0.002^{\rm s}$ 
 	&59$^\circ 25'14.78'' \pm 0.01''$		
	&$8.8^\circ$
	&---
	&0.12\,Jy\ \ \ \ \ \ \ \ \ 0.20\,Jy 
	&5		
\\
1300+580
	&13$^{\rm h}$02$^{\rm m}$52.465$^{\rm s} \pm 0.002^{\rm s}$ 
 	&57$^\circ 48'37.62'' \pm 0.01''$		
	&$5.5^\circ$
	&---
	&0.13\,Jy\ \ \ \ \ \ \ \ \ 0.20\,Jy 
	&5
\\
\hline
\multicolumn{7}{l}{{\it Position refs.}: 
(1) This paper; (2) Williams et al. (1996);
(3) Richards et al. (1998); (4) Aussel et al. (1999); 
}  
\\
\multicolumn{7}{l}{(5) Patnaik et al. (1992).  {\it Redshift refs.}: 
(6) Rowan-Robinson (1999);  (7) Fern\'andez-Soto et al. (1999);  
(8) Zepf et al. (1997)
}  
\\
\end{tabular}
\end{table*}

Richards (1999) argued against high redshifts ($z>2$) 
for the HDF sub-mm sources, 
from plausible associations of
the SCUBA sources with nearby radio sources detected in his VLA  survey at
20\,cm (1.4\,GHz) and the  identification of some of these non-thermal radio
continuum  sources  with optical galaxies in the Hubble Deep Field. From the
radio luminosity function of starburst galaxies, Richards argued  that if the
submm dust sources had cm-radio detections, then they  must be either $z<2$
star-forming galaxies or, if at higher redshift, heated by active galactic
nuclei (AGN).  To clarify these issues, 
 it is important to obtain good positions for
the dust sources by millimeter interferometry to examine possible 
coincidences with the Hubble Deep Field optical galaxies and 
VLA radio sources.

The strongest object in the SCUBA survey is the source HDF 850.1, which has a
flux density of 7.0$\pm 0.4$\,mJy at 850\,$\mu$m.     Follow-up photometry of
HDF 850.1 by Hughes et al. (1998)  
yielded a flux density of $2.1\pm 0.5$\,mJy at a wavelength
of 1.35\,mm, confirming that the continuum   is optically thin emission by dust.
At longer millimeter wavelengths, the Hubble Deep Field is known to be empty of
sources.  With the BIMA interferometer, 
Wilner \& Wright (1997) found no sources at 2.8\,mm,  to a
5$\sigma$-limit of 3.5\,mJy. Because the 1.3\,mm flux 
density of HDF 850.1 is close to
the minimum detectable level in a 10-hr integration at this wavelength,  
this strongest source in the
SCUBA survey may be the {\it only} source in the Hubble Deep Field that can be
detected at present in reasonable observing times 
with millimeter interferometers. To measure more accurately
the position of this source, we observed it at 1.3\,mm and 3\,mm with the IRAM
Interferometer on Plateau de Bure, France.

\begin{figure}
\vspace{0.3cm}
\psfig{figure=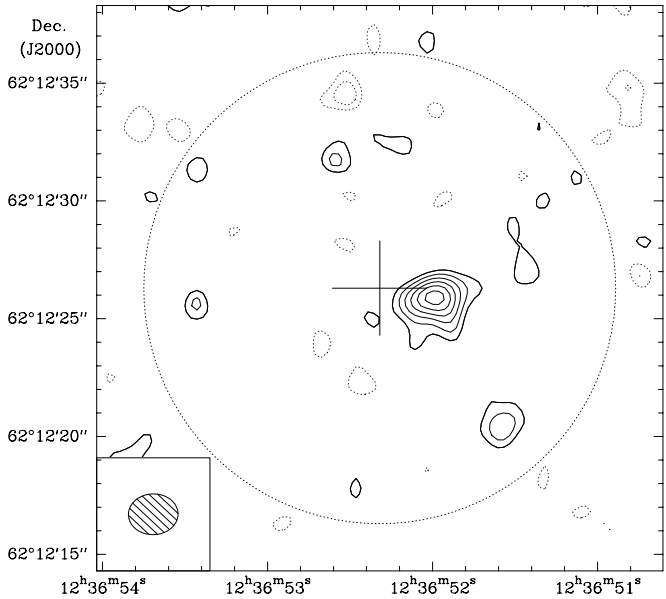,width=8.7cm,angle=-90 }
\caption[1mm map]
{IRAM interferometer 
map at 1.3\,mm (236.3\,GHz) of the continuum emission in a 24$''$ field centered
on the SCUBA position of the submm source HDF850.1 (cross; the arms of the
cross correspond to the 3$\sigma$  positional uncertainty of Hughes et al.
1998). The synthesized beam is 2.1$'' \times 1.7''$ (box at lower left). The map
has not been corrected for the response of the  primary beam of the 15\,m dishes
at  1.3\,mm  (20$''$ FWHM; dashed circle). 
The first positive  and negative (dashed) contours are at $\pm
0.5$\,mJy (2\,$\sigma$), and the contours thereafter are  in steps of 0.25\,mJy
(1\,$\sigma$). 
}\end{figure}

\begin{figure}
\vspace{0.3cm}
\psfig{figure=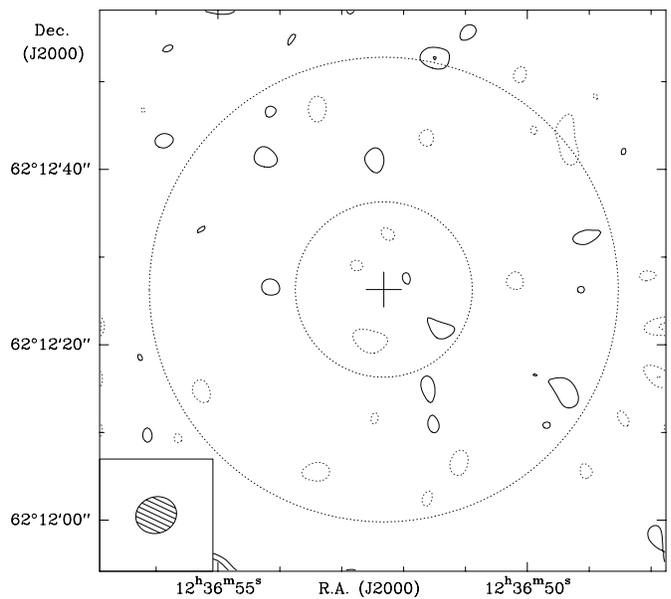,width=8.7cm,angle=-90 }
\caption[3mm map]{
Interferometer 
map at 3.4\,mm (88.7\,GHz) of the continuum emission in a 64$''$ field centered
on the original SCUBA position (cross) of the submm source HDF850.1.
No source is detected, to a limit of 0.36\,mJy (3$\sigma$).  
The synthesized beam is 4.7$'' \times 4.2''$ (lower left).
The map has not been corrected for the response of the primary beam 
at 3.4\,mm (53$''$ FWHM; outer dashed circle). 
The inner dashed circle shows the 
primary beam at 1.3\,mm (see Fig.~1).
The first positive and 
negative (dashed) contours are at $\pm 0.24$\,mJy (2\,$\sigma$), 
and the contours thereafter are in steps of 0.12\,mJy  
(1\,$\sigma$). 
}\end{figure}
\section{Observations}
The observations were made at 1.3\,mm and 3\,mm simultaneously, 
in the interferometer's  compact
configuration D on November 17 and 22, 1998, and in the  more 
extended configuration C2
on December 15, 16, and 17, 1998.  The total integration time was about 40
hours, in excellent weather,  with precipitable water vapor content $\approx$
1.5\,mm, and r.m.s.\ phase errors at 1.3\,mm of $\leq 30^\circ$.  
The five 15\,m
dishes give 10 interferometer baselines  from 24\,m to 80\,m in configuration D,
and 24 to 183\,m in configuration C2.   At 1.3\,mm, the observing frequency was
236.3\,GHz, and data were taken in upper and lower sidebands separated by 
3\,GHz.  At
3\,mm, data were taken in lower sideband only, at 2.8\,mm 
(105.7\,GHz) on November 17, and at 3.4\,mm 
(88.7\,GHz) on all the other dates. The SIS receivers had 
equivalent system temperatures outside the atmosphere 
of 150\,K SSB at 3\,mm, and  250 to 400\,K SSB at 1.3\,mm. The spectral
correlators covered 954\,\kms \ at 3.4\,mm and  670\,\kms \ at 1.3\,mm, with
resolutions of 8 and 4\,\kms , respectively.  Flux densities were 
calibrated with
3C273 and MWC349, for which we adopted the following values:
for 3C273,  11\,Jy at 1.3\,mm and 18\,Jy at 3.4\,mm; 
for MWC349, 1.7\,Jy at 1.3\,mm and 0.9\,Jy at 3.4\,mm.  The systematic 
uncertainties in the flux scales are estimated to be $\pm 5$\% at 3.4\,mm 
and $\pm 10$\% at 1.3\,mm.

 Phases were  calibrated with the radio sources 1044+719,
1125+596, and 1300+580, for which we adopted the positions listed in 
{\bf Table~1}.  All three phase calibrators were observed every 20\,min, at 
1.3\,mm and 3\,mm simultaneously, and the data were 
phase calibrated with a weighting 
 by the square of their signal-to-noise
ratios and by the inverse square of the system
temperature. 
To correct the amplitudes and the phases 
at both 3\,mm and 1.3\,mm for atmospheric seeing, 
we used  the 1.3\,mm continuum
total power measurements to compute the changes in 
electrical path length due
to short-term fluctuations in the atmospheric water vapour content.
For the 1.3\,mm data, we then adopted the phase vs.\ time 
calibration curve at 3\,mm
(scaled to 1.3\,mm) and on top of this, solved for a second-order 
phase calibration curve to fit the phase residuals at 1.3\,mm.

\begin{figure}
\psfig{figure=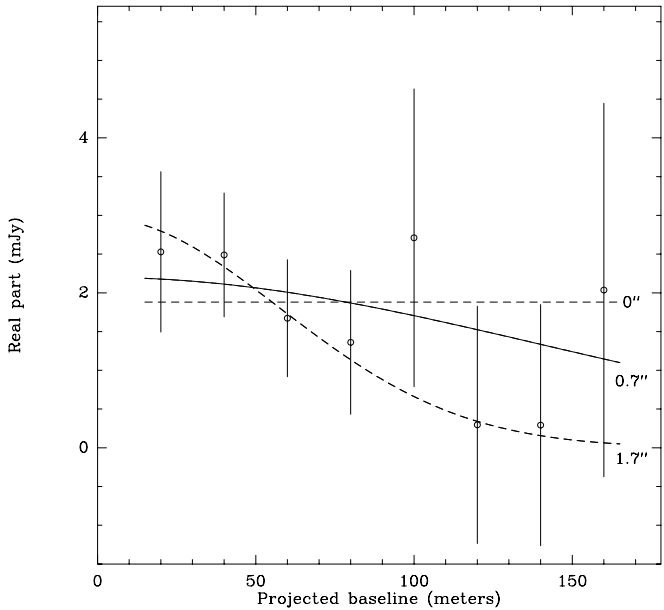,width=8.7cm,angle=-90 }
\caption[Visibility plot at 1.3\,mm]{
Visibility plot for the continuum source detected at 1.3\,mm. The plot shows the
real part of the visibility amplitude  vs.\ baseline, for $u,v$-plane data
averaged in  circular bins of 20\,m spacing, with 1$\sigma$ error bars.    The
data do not have a high enough signal-to-noise ratio to yield an angular size.
The ``best-fit" (solid curve) is for a 2.2\,mJy gaussian source of FWHP diameter
0.7$''$.  The dashed curves show the visibility functions at the 2\,$\sigma$ 
uncertainty
 of  the fit, for a 1.7$''$ source (lower dashed curve)  and a point source
(upper dashed curve).
}\end{figure}
 We used both uniform and natural weighting of the 
$u,v$ plane data to make maps.  There is not much difference in the final result, 
and the maps shown here 
({\bf Figs.~1 \& 2}) are the ones made with natural weighting.
The 1.3\,mm map in {\bf Fig.~1} includes all the data, 
in both sidebands and in the compact and extended configurations,
and has been deconvolved with the CLEAN algorithm by Clark (1980), 
with 100 iterations.  

\section{Results}
\subsection{Detection at 1.3\,mm}
A continuum source was well detected at 1.3\,mm (236.3\,GHz),  
in the D configuration on November 17 and November 22 in the double sideband 
data, and in the
upper and lower sidebands separately.  All of these data subsets 
 agree in the source position and flux density, 
within the errors.  On the map that
results from adding all the data together ({\bf Fig.~1}),  the flux density is
$2.2\pm 0.3$\,mJy. This result agrees with the single-dish value 
of $2.1\pm
0.5$\,mJy measured by Hughes et al.  (1998) at 1.35\,mm.

The 1.3\,mm source is 
at a position  $2.4''$ west and 0.7$''$ south of the SCUBA position
of the submm source HDF 850.1 found by Hughes et al. (1998). 
The SCUBA detection was made with the 15\,m JCMT single dish, with a 14.7$''$
beam at 850\,$\mu$m.  Hughes et al.\ quoted r.m.s.\ positional 
uncertainties of 0.7$''$ in each coordinate.
The interferometer position differs from the SCUBA position at the 
3.6\,$\sigma$ level, indicating that there may have been an unsuspected
systematic error in the SCUBA positions, which in turn may affect some
of the other submm/optical associations suggested by Hughes et al. (1998).

\subsection{Size of the 1.3\,mm source}
The source is too weak to measure an angular size. 
 In the maps from the longer-baseline C2
configuration alone, the flux at the source position is $1.3\pm 0.5$\,mJy, 
so the source may be partially resolved.
 Our fits to the data in the $u,v$-plane yield a ``best fit" FWHP diameter of 
$0.7''\pm 0.5''$.  At the 2\,$\sigma$ level, 
the visibility data ({\bf Fig.~3}) agree
with sizes ranging from 1.7$''$ to a point source.   
What size should we expect?  Since we observe mm/submm dust radiation,
the source cannot be very hot.  For example,
a circular source at $z\approx 1.7$ (see below)
\footnote{We use $H_0 = 50$\,\kms\,Mpc$^{-1}$ and $q_0 = 0.5$ 
in this paper.} 
with a dust temperature of 35\,K, which has a dust opacity of unity 
at an emitted
wavelength of 100$\mu$m, would yield a flux of 2.2\,mJy at an observed
wavelength of 1.3\,mm if its diameter were 0.3$''$.
  A less opaque source would be even larger.  

\subsection{No detection of 3\,mm continuum}
No source was detected at either 2.8\,mm  (105.7\,GHz) or at 3.4\,mm
(88.7\,GHz), to $3\sigma$ limits of 0.5\,mJy and 0.4\,mJy respectively 
({\bf Fig.~2}).
This is consistent with the 1.3\,mm continuum being optically thin emission by
dust. For a typical dust continuum  spectral index of 3.5, the expected fluxes
of HDF 850.1  would be only 0.13\,mJy 
at 2.8\,mm and 0.07\,mJy at 3.4\,mm.  

\subsection{No spectral lines detected}
Although our main goal was the detection of  dust emission at 1.3\,mm, we chose
the frequency in the upper sideband, 237.824\,GHz, to coincide with the
CO(9--8) line at $z$ = 3.36, the tentative redshift assigned by Zepf, Moustakas,
\& Davis (1997) to the optical source 3-577.0 located 1.4$''$  
east of the  $z\approx 1$ elliptical 
galaxy 3-586.0.   
No line was detected in the upper sideband, to a limit of 6\,mJy beam$^{-1}$ in
channels 32\,\kms \ wide in the velocity range $-$300 to  +367\,\kms\ relative to
the frequency of the CO(9--8) line at this  redshift.  We also detected no line
in the lower sideband, 3\,GHz lower in frequency.

At 3\,mm, we initially chose 105.7\,GHz to search for the CO(4--3) line at $z$ =
3.36.  At this frequency, no line was seen, to a limit  of 2\,mJy\,beam$^{-1}$
in channels 30\,\kms \ wide,   in the velocity range $\pm 400$\,\kms\  around
the frequency of the CO(4--3) line at this redshift. Later on, we tuned the
3\,mm receivers to 88.738\,GHz  to search for the CO(1--0) line at  $z$ = 0.299,
the redshift of the bright disk galaxy, 3-659.1,  located 10$''$ to the
southwest of HDF 850.1.  In velocity channels 34\,\kms \ wide,  no line was seen
to a limit of 2\,mJy beam$^{-1}$, anywhere in our 53$''$ primary beam, in the
velocity range $-$450 to +460\,\kms \ centered on 88.738\,GHz.  For a galaxy
with lines $\sim 200$\,\kms \ wide, this corresponds to a limit of 0.4\,Jy
\kms\  for the \CO10 line flux.   For comparison, the Milky  Way has a 
\CO10 line luminosity of $4\times 10^8$\,K\,\kms \,pc$^2$ inside the 
solar circle (Rivolo \& Solomon 1988), so 
at $z\approx 0.3$, the
Milky Way would have a \CO10 line flux of 0.05\,Jy\,\kms .

\section{Astrometry}

We estimate the r.m.s.\ positional uncertainty
  to be 0.3$''$ for the  centroid of the
1.3\,mm dust source.  This limit is set by the baseline accuracy of
the interferometer,  the availability of phase calibrators in this part of the
sky, the  signal to noise ratio, and the seeing.  Thus far, 
the best astrometry done with the
interferometer was for the source W3(OH), with phases
calibrated on the quasar 0224+671, located 5$^\circ$ away.  In those projects
(Wink et al. 1994;  Wyrowski et al. 1997; 1999),  the global statistical
positional repeatability was 0.05$''$ r.m.s.,  including the systematic errors 
in  the baseline geometry.   To this must be added the $0.01''$ uncertainty in
the Jodrell Bank--VLA  position of the calibrator source.   For
the current Hubble Deep Field project,  our strongest 
calibrator is 14$^\circ$ away ({\bf Table~1}), so
the r.m.s.\  error in position  would be three times worse, namely $0.2''$,
also with Jodrell Bank-VLA calibrator uncertainties of 0.01$''$ 
(Patnaik et al. 1992).  

\begin{figure*}
\vspace{0.6cm}
\psfig{figure=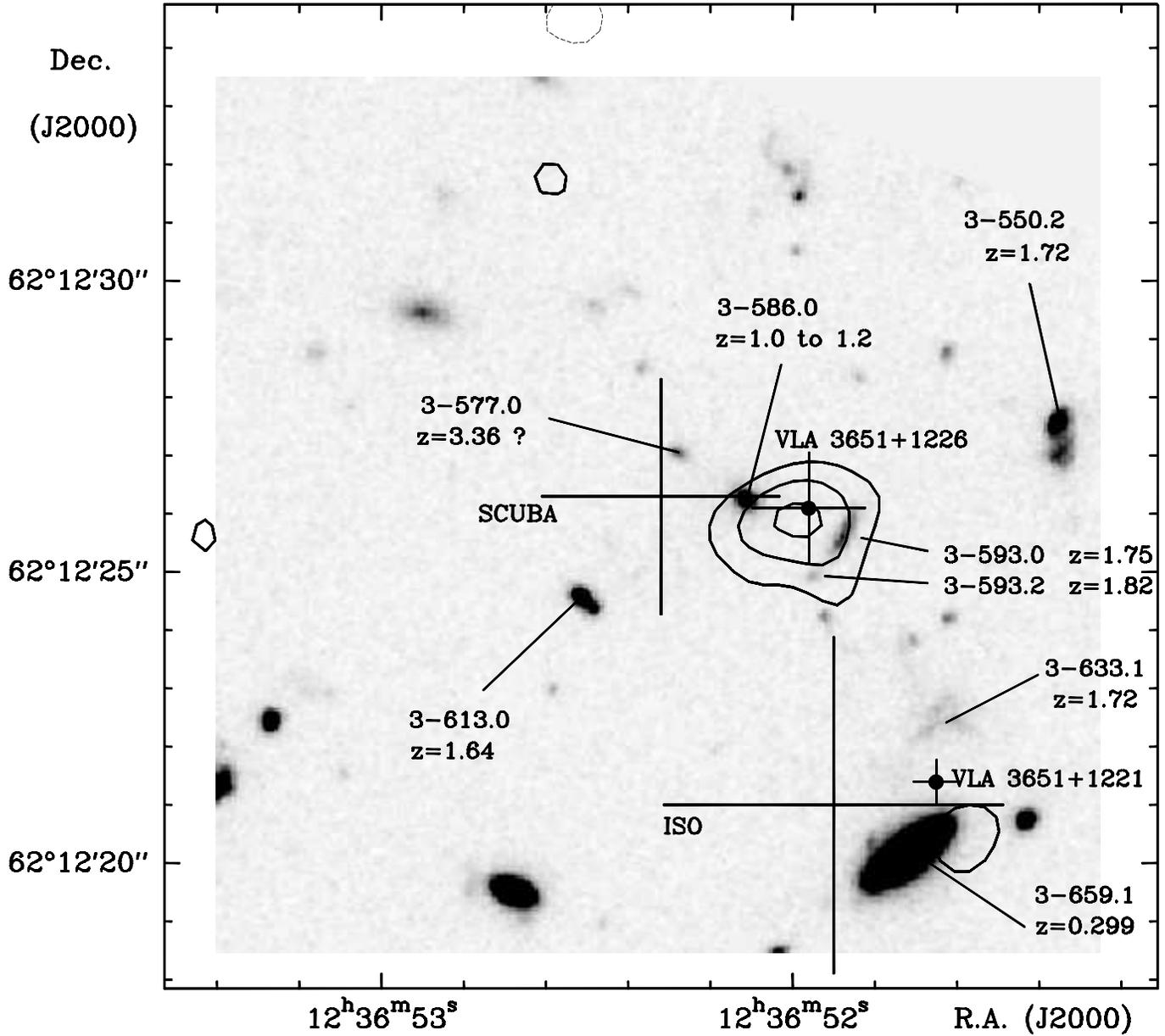,width=18.0cm,angle=-90 }
\caption[1mm map]
{IRAM interferometer 
map at 1.3\,mm (236.3\,GHz) of the dust continuum emission superimposed on 
a greyscale version of the {\it BVI} image from the Hubble Deep Field.
The map is the same as in Fig.~1, but here we only plot the 1.3\,mm contours
at 0.75\,mJy (3\,$\sigma$), 1.25\,mJy (5\,$\sigma$), and 1.75 mJy
(7\,$\sigma$).  The cross marked SCUBA indicates the 
position and $3\sigma$ uncertainty of HDF 850.1 as given by Hughes et al.
(1998).  Small crosses with black dots indicate the positions 
and $1\sigma$ uncertainties of the
VLA sources (Richards 1999) and the large cross marked ISO 
indicates the position and $1\sigma$  uncertainty of the ISOCAM 15$\mu$m
source (Aussel et al. 1999).
 Optical objects are identified with their 
{\it photometric} redshifts, from the references in {\bf Table~1}, except for  
the optical source 3-577.0, which 
has a tentative {\it spectroscopic} redshift
of 3.36 (Zepf et al. 1997).
}
\end{figure*}

To reduce errors, we chose the three phase calibrators at earlier and  later
right ascensions, and lower and higher declinations than the  Hubble Deep Field
({\bf Table~1}). This should formally reduce the noise in the baseline 
errors and in 
our weighted average of  calibrators  by a
factor $\sim \sqrt{2}$ to $\sqrt{3}$.  
To check internal
consistency of the reduction software, we made maps of the calibrator sources,
and on the maps, the position errors were $< 0.1''$.  
The uncertainties in their  measured
cm-radio positions from the Jodrell-VLA survey are ten times lower ({\bf
Table~1}).  Two of the calibrators are weak, $\sim 0.1$\,Jy at 1.3\,mm, but
their signal-to-noise ratio is quite satisfactory at 3\,mm.  
Because we subtract the scaled 3\,mm phase fluctuations from those at 1.3\,mm, 
we can improve the fit of the phase solution at 1.3\,mm, but there  
remains a residual
instrumental phase error, which brings the astrometric part of the 
position error  to  0.2$''$ r.m.s.  

In addition to the astrometric error, there is a statistical error related 
to the signal-to-noise ratio. 
For a point source, this position error due to noise is 
\begin{equation}
  \Delta \theta \approx  (B/2) / (S/N)  
\end{equation}
where $B$ is the synthesized beam, and $S/N$ is the signal-to-noise ratio. For
our map in configurations D + C2, the beam $B$ is $2.1''$, and the
signal-to-noise ratio is   $S/N$ = 7, so the position error due to noise is
$\Delta \theta = 0.2''$ r.m.s. The convolution of  the astrometric error
(0.2$''$),  and the noise error (0.2$''$) then gives a root sum square error of
0.3$''$.

\section{Source Identification} 
The identification of the 1.3\,mm dust source with an optical galaxy  or a
cm-radio source in  the Hubble Deep Field depends strongly on the accuracy of
the  measured positions.  The radio-optical registration of the Hubble Deep
Field is  described by Williams et al. (1996).  Their estimated accuracy of this
absolute registration is 0.4$''$.   The estimated accuracy of the  1.3\,mm IRAM
position is 0.3$''$.  

{\it Radio:} 
The new 1.3\,mm position clearly rules out the  identification with the optical
source  3-577.0 proposed by Hughes et al. (1998) from the SCUBA data. It 
also rules out the identification, suggested by Richards (1999), of the 
1.3\,mm dust source with the 
VLA source, 3651+1221, located  4.9$''\pm 0.4''$ to the southwest
(see further discussion of this source in the Appendix).
The 1.3\,mm dust source coincides with the tentative VLA source 3651+1226,
which is a $4.5\sigma$ detection at 3\,cm (Richards et al. 1998).

{\it Optical:} On the Hubble Deep Field image ({\bf Fig.~4}), 
the closest object to the  1.3\,mm dust source
is the arc-like feature comprising the optical sources 3-593.0 and 3-593.2.   
From the optical coordinates
published by Williams et al. (1996), this arc-like feature 
is  0.8$''\pm 0.7''$  southwest of
the 1.3\,mm dust source, where the $0.7''$  error is the 
sum of the
optical and 1.3\,mm position errors  ({\bf Table~1}). 
We must take the sum of the errors rather than the root sum square 
because the radio-optical registration uncertainty is a systematic error
rather than a random error.

\begin{figure}
\psfig{figure=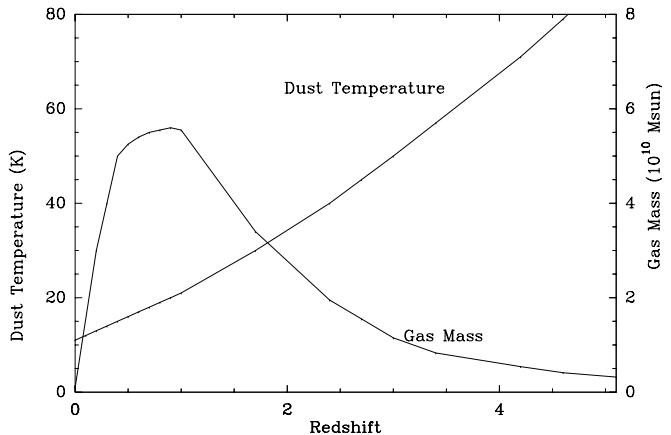,width=8.7cm,angle=-90 }
\caption[Allowed values of dust temp. and gas mass]{
Allowed values of the dust temperature and gas mass, as functions of  redshift,
for an unlensed source, 
that are consistent with the observed 1.3\,mm flux of 2\,mJy and the
450$\mu$m/1300$\mu$m flux  ratio.  ``Low" redshift ($z\approx 1$) solutions
require low temperature (20~K) and high gas mass ($6\times 10^{10}$\,\Msun).
``High" redshift ($z\approx 3$) solutions can have  higher dust temperatures
(60~K) and lower gas mass ($8\times 10^{9}$\,\Msun).  
}
\end{figure}

{\it The Elliptical:} 
The next nearest optical object is the source 3-586.0,  
located 1.0$''\pm 0.7''$ east of
the 1.3\,mm dust source.   The optical source 3-586.0 is an  elliptical galaxy
at  a photometric redshift of $1.0\leq z \leq 1.2$ (Hogg et al. 1996;  Lanzetta
et al. 1996 (their source number 3-306); Mobasher et al. 1996; Gwyn \& Hartwick
1996;  Sawicki et al. 1997; Wang et al. 1998;  Fern\'andez-Soto et al. 1999 ---
their source 303). The optical photometry of this elliptical galaxy  is
tabulated by Fasano et al. (1998),  and its  spectral energy distribution (SED)
is given by Cowie (1999).  The spectrum rises from the optical regime to a
maximum near 2\,$\mu$m, where it flattens out at a level of $\sim 10$\,$\mu$Jy,
consistent with the  non-detection at a level of 23\,$\mu$Jy at  15\,$\mu$m
(Oliver et al, in prep., reported by Hughes et al. 1998; Aussel et al. 1999).

From the position alone, the dust source  could be the arc-like feature 
or the elliptical.  We argue,
however, that the 1.3\,mm dust source is probably associated with the arc-like
feature.   Although we cannot associate the dust source with this optical 
object from the astrometry, we now  present several physical  arguments  that
support this association and disfavor the elliptical.

\subsection{Dust flux ratios, dust temperature, and gas mass} 
The source HDF 850.1 
is not detected at 450\,$\mu$m to a limit of 21\,mJy (Hughes et al.
1998), nor is it in the IRAS Faint Source Catalog, nor in the ISO images of the
Hubble Deep Field  at 6.7\,$\mu$m and 15\,$\mu$m (Aussel et al. 1999).   
We thus 
have a limit on the 450\,$\mu$m to 1300\,$\mu$m  flux ratio,
where  the emission is optically thin  (the spectral index is $\sim 3$).  
This ratio is $S(450\mu$m)/$S(1300\mu$m) $\leq$ 10.  Let
the corresponding observed frequencies (667 and 236\,GHz) be $\nu_2$ and
$\nu_1$, respectively. Because the shape of the Planck function is preserved
with redshift,  the ratio is the same at the emitted frequencies, namely  
\begin{equation}
{ { S(\nu_2) } \over { S(\nu_1) } }
= \bigg( { {\nu_2} \over {\nu_1} }\bigg)^3
{
{(e^{h\nu_1 (1+z)/kT_d} -1)} 
\over
{(e^{h\nu_2 (1+z)/kT_d} -1)} 
}
{
{(1-e^{-\tau(\nu_2 (1+z))})}
\over
{(1-e^{-\tau(\nu_1 (1+z))})}
}
\end{equation}
where $T_d$ is the dust temperature and 
$\tau(\nu)$ is the dust optical depth. Experience in millimeter detection
of dust at high redshifts during the  past decade shows that sources are only
detected when they contain large amounts of dust ($\sim 10^8$\,\Msun )
--- as in the IR ultraluminous
galaxies which have dust opacities equal to unity near a 
wavelength of
100\,$\mu$m (e.g., Downes, Solomon, \& Radford 1993).  At this wavelength 
 (frequency 3\,THz),  one may approximate the dust
opacity as   $\tau(\nu) = (\nu/3)^n$, where $\nu$ is the emitted frequency in
THz, and  $n$ is the dust emissivity index, where $n\approx 1.5$ to 2.
Regardless of whether the dust is optically thick or thin, however, at ``low"
redshifts ($z<1$), the ratio in  eq.(2) can be as low as the observed
value of $\leq 10$  only with dust temperatures  $T_d \leq 20$\,K. {\bf
(Fig.~5)}.   

From the mere detection of dust at 1.3\,mm with a flux density of 2.2\,mJy, we
can make a useful deduction about the gas mass from the formula 
(cf. Downes et al. 1992)
\begin{equation}
{ M  \over { [{\rm M}_\odot] }  }  =  
{ {1.6\,10^6}\over  {(1+z)} }
{ S_{\nu_{\rm obs}} \over {\rm [Jy]} } 
\bigg[{ D_L \over {\rm [Mpc}] }\bigg]^2
\bigg[{ T_d \over {\rm [K]} }\bigg]^{-1} 
\bigg[ { \nu_{\rm em} \over {\rm [THz]} }  \bigg]^{-(2+n)} 
\end{equation}
where $M$ is the gas mass,  $S$ is the flux density, 
$D_L$  is the luminosity distance, 
$T_d$ is the  dust temperature,  $\nu_{\rm em}$ is the emitted frequency, 
and $n$ is the dust emissivity index, which we adopt to be $n=1.5$.
The main assumption is that the emission comes from optically thin dust. 
The equation is for 
a dust mass absorption coefficient of 0.11\,m$^2$\,kg$^{-1}$ of dust at 
an emitted wavelength of 1.3\,mm, as in dense molecular clouds 
(e.g., Kr\"ugel \& Siebenmorgen 1994, their Fig.~12), and a 
gas-to-dust mass ratio of 100.  
The uncertainties are about a factor of three in 
each direction --- higher and lower  (for an analysis, see Hughes, Dunlop,
\& Rawlings 1997).   Higher gas-to-dust ratios mean even larger gas masses.  

To have a flux density 
of 2\,mJy at 1.3\,mm with the low dust temperature derived above,
a  ``low''-redshift ($z\approx 1$) source would need a very 
high molecular gas mass,  approaching $10^{11}$\,\Msun . 
Lower dust temperatures
or less opaque dust or higher gas-to-dust ratios would 
all imply even higher gas masses. Such an enormous quantity of dust and
molecular gas would have dramatic consequences for star formation, and such a
large, gas-rich galaxy at low redshift would have been easily seen on the HDF
images or in the follow-up optical spectroscopy of these sources (e.g. Zepf et
al. 1997).    For comparison, a large galaxy like NGC~891, at a distance of
10\,Mpc,  has a total 1.3\,mm dust continuum flux of 730\,mJy, corresponding to
a molecular gas  mass of $1.5\times 10^9$\,\Msun\ (Gu\'elin et al. 1993), about
the same as that of the Milky Way.    If NGC~891 or the Milky Way were moved to
$z=1$,  at an emission distance (angular size distance) of  1.75\,Gpc, then the
1.3\,mm observed flux  would be only 36\,$\mu$Jy, that is, 60 times lower than
the dust flux we detect.

At higher redshifts, the observed 
limit on the 450\,$\mu$m to 1300\,$\mu$m  flux ratio
 allows the dust temperature to be
higher.  For example, at $z = 3$, the dust temperature can be 60\,K, comparable
with the {\it blackbody} dust temperatures deduced for IR ultraluminous galaxies from
their IRAS fluxes (e.g. Downes \& Solomon 1998).     The
corresponding molecular gas masses deduced from the observed 1.3\,mm flux also
become plausible --- a few $\times 10^9$\,\Msun ,  even in the absence of
gravitational magnification. Because the line of sight to the source passes so 
close to the elliptical --- a good gravitational lens---
 the observed mm and submm fluxes may be 
gravitationally magnified, and the real gas mass even lower.

In summary, our argument,  based on the mm and submm fluxes alone, that the dust
source cannot be  a ``low" redshift ($z \leq 1$) object,  is  as follows: 
{\it a)} the optical depth is not high;  
{\it b)} the observed flux ratio may  be obtained at
low redshift only if the dust temperature is sufficiently low ($<20$\,K), and
{\it c)} the observed flux plus a low temperature requires an excessive gas mass.  

\subsection{Other arguments against identifying HDF 850.1 with the elliptical 
galaxy 3-586.0}
A number of other facts  argue against the optical source 3-586.0, an elliptical
galaxy at $z \approx 1$, being the optical object associated with the 1.3\,mm
dust source.

{\it 1)}  The gas and dust masses of ellipticals  are 
generally too low for 1.3\,mm dust detections 
($M_{\rm gas} < 10^{10}$\,\Msun ).    Even in nearby ellipticals, there are
very few detections (see, e.g., the review by Knapp 1999).   
The three galaxies with dust 
detections $>3\sigma$ reported by Wiklind \& Henkel (1995) have 
gas masses ranging from 
$1\times 10^5$ to $7\times 10^8$\,\Msun .
The  much larger dust and gas mass  predicted by the physical arguments in 
the previous section would  be totally inconsistent with the gas
content of an elliptical galaxy like 3-586.0.   
Such galaxies typically have a total dynamical mass (mostly stars)
of $3\times 10^{10}$\,
\Msun \ within a radius of 1\,kpc, and, with a de Vaucouleurs profile,
 enclose a total mass of $2\times 10^{11}$\,\Msun \ at a radius of 7\,kpc 
( 0.85$''$ at $z\approx 1$).  
This argument from
the dust and gas mass needed to yield 2\,mJy at 1.3\,mm 
is equivalent to the empirical luminosity argument  by  Hughes et al. (1998) 
--- that  
the mm/sub-mm luminosity of the source HDF 850.1 is two orders of magnitude  too
high to be emission by dust in the $z\approx 1$  elliptical galaxy 3-586.0.
Therefore the dust source cannot be the  $z=1$ elliptical galaxy.

{\it 2)} Dust has not yet been detected at 1.3\,mm in any  ``normal '' galaxy at
$z\approx 1$.   The reasons are firstly, that  the   redshift is not high enough
to shift warm dust emission into the millimeter band, and secondly, that the
emission distance (angular size distance)  is close to its maximum.    A
striking exception is the extremely red,  $z=1.44$ galaxy  HR~10 (or ERO
J164502+4626.4), for which  Cimatti et al. (1998) and Dey et al. (1999) find
mm/submm dust fluxes  corresponding to  an astonishing far-IR luminosity of
$\sim 10^{13}$\,\Lsun , a  molecular gas mass of $\sim 10^{11}$\,\Msun , and  an
apparent star formation rate of 1000 to 2000\,\Msun \,yr$^{-1}$.    It will be
important to  investigate whether this object is gravitationally lensed or not.

{\it 3)}  We argued above that if the object is at $z\approx 1$, the dust  must
be cold.   So far, however,  the mm detections of dust in high-$z$  objects  are
mainly of ``warm" dust ($T > 30$\,K), probably  circumnuclear regions heated by
an AGN and/or a nuclear starburst.   Therefore it is most unlikely that we are
detecting the elliptical at $z\sim 1$.

\section{Discussion}
\subsection{The dust source HDF850.1 as an unlensed source}
Hughes et al. (1998) gave a very plausible interpretation of HDF850.1 as a
high-redshift starburst galaxy,  similar to the ultraluminous infrared galaxies
at $z<0.3$. As discussed above,  from the interferometer
position measurement and the physical arguments about the gas mass,  it is
likely that the dust source HDF850.1 is associated with  the optical arc-like
feature 3-593.0/593.2. The photometric redshift estimates  of the arc are $z=
1.73$ (Rowan-Robinson 1999) and   $z=1.76$ (Fern\'andez-Soto et al. 1999).   In
this part of the sky there seems to be a  cluster or a group of galaxies at
photometric redshifts  $z\approx 1.7$ (the objects  3-593.0, 3-633.1,  3-550.2,
and 3-613.0). It is conceivable that the dust source is at this redshift as
well,  and that it is part of the same object as the optical  arc-like feature.
Such a redshift for HDF850.1 would be well within the redshift distribution 
of the identified sources detected so far with the SCUBA array in other fields 
(e.g., Barger et al. 1999a,b; Eales et al. 1999; Lilly et al. 1999).

No optical lines were detected in the arc by  Zepf et al. (1997).   Why not?  A
simple explanation is that the object  is at $1.2 < z < 2.4$, consistent with
the photometric redshifts,  so that no strong lines are shifted into the
optical.  As shown by the data of Zepf et al. (1997),  the arc  3-593.0 is not
likely to be  at $z\geq 3$, because  it is too blue --- it shows up nicely in
the HST F450W band, and it has significant flux in the F300W band.  It is not a
``U-band dropout",  that is, the hydrogen along the line of sight does not
greatly decrease the flux in the F300W band, as it would if the object were at
$z\geq 3$, again consistent with  the photometric redshift of  $z\sim 1.7$. 
 The arc is much bluer than, for example, 
the object 3-577.0, which has a photometric redshift of 2.88 (see, e.g, 
the colors tabulated in Zepf et al. 1997).

\begin{figure}
\psfig{figure=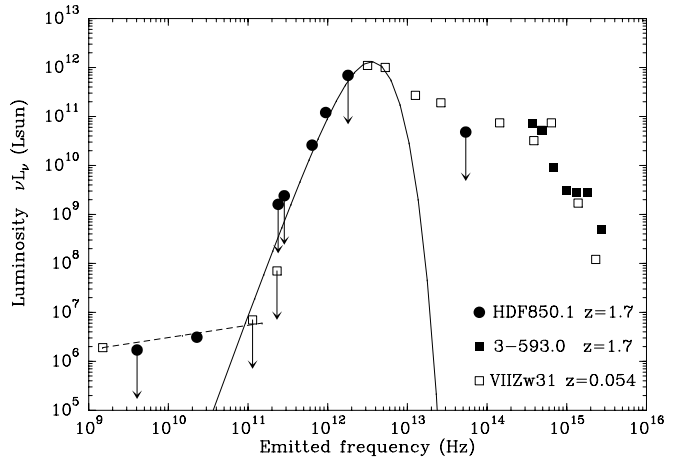,width=8.7cm,angle=-90 }
\caption[SED for z=1.7]{
Spectral energy distribution for the dust source HDF 850.1 and 
the optical source 3-593.0 if they are at  
 $z\approx 1.7$, compared with the spectral energy 
distribution of the ultraluminous IR galaxy VII~Zw~31 at 
$z=0.054$.  Data for HDF 850.1 are from 
this paper and Hughes et al. (1998). For 3-593.0, fluxes are from 
Fern\'andez-Soto et al. (1999). For the VLA source
3651+1226, cm-radio fluxes are from Richards et al. (1998) and Richards (1999).
For VII~Zw~31, IR and optical data are from Trentham et al. (1999); mm limits
are from Downes \& Solomon (1998), and the 20\,cm flux is from 
Condon et al.\  (1996).  The solid curve shows a spectrum of
the far-IR emitting dust component, from the model described in the text.
The dashed line is for an assumed synchrotron spectral index $\alpha = -0.75$
for VII~Zw~31.
}
\end{figure}

In fact, the data on HDF 850.1, if combined with the optical fluxes of the
optical arc-like feature 3-593.0 and a redshift of 1.7,  agree rather well with
the spectral energy distribution of the ultraluminous infrared  galaxy
VII~Zwicky~31.  This object, which was interpreted by Djorgovski et al. (1990)
as a merger-induced starburst galaxy,  was recently selected for study with the
HST  by Trentham, Kormendy, \& Sanders (1999) because its  ``cool'' spectral
energy distribution indicates it is an  ultraluminous galaxy that derives its IR
luminosity from an embedded starburst only,  with no ``warm'' AGN component.
{\bf Figure~6} shows this comparison with VII~Zw~31, if    HDF 850.1 is
associated with the optical object 3-593.0, that has a  photometric redshift of
$z\approx 1.7$. For VII~Zw~31, we took the IRAS and HST data from Trentham et
al. (1999), re-scaled for $H_0 =50$\,\kms \,Mpc$^{-1}$. The luminosities of the
far-IR peak of HDF 850.1 and the optical emission of 3-593.0,  calculated for $z
= 1.7$,  correspond quite well to the observed luminosity of VII~Zw~31.  
The solid curve in {\bf Fig.~6} shows the predicted spectrum of the 
cool dust component from the radiative-transfer model for VII~Zw~31 by
Downes \& Solomon (1998).  In that model, the dust temperature varies with
galactocentric radius. The denser dust that is opaque at 100\,$\mu$m 
has a temperature of 50\,K.  The millimeter dust emission is 
transparent over the full extent of the source, for which the global spatial
average of the dust temperature is $\sim 35$\,K.

There is also a good agreement  in the size of 3-593.0 and the dust region in
VII~Zw~31.    On the Hubble Deep Field image,  the bright part of 3-593.0 has a
radius  of 0.22$''$, or $\sim 1.9$\,kpc at $z\approx 1.7$. The total emission
along the arclike feature, including  the faint optical object 3-593.2, has an
radius of 0.67$''$, or 5.6\,kpc. In VII~Zw~31, the relevant dimensions are
nearly the same. The half-power radius of the rotating molecular gas disk in
VII~Zw~31 is 1.7\,kpc,  with an outer radius of 5\,kpc (Downes \& Solomon 1998),
and  the ring of star-forming knots in VII~Zw~31 has a radius of 3.4\,kpc
(Mazzarella et al. 1999; image in Trentham et al. 1999). 
This good match to the star-forming region of VII~Zw~31 in both size and
spectral energy distribution, if HDF 850.1 is at $z\approx 1.7$ and associated
with 3-593.0, suggests a  plausible interpretation of HDF 850.1 as an
ultraluminous IR galaxy with the same intrinsic 8\,$\mu$m to 1000\,$\mu$m
luminosity as VII~Zw~31,  namely, $L_{\rm IR}$ = $2\times 10^{12}$\,\Lsun .

Although the best-fit photometric redshift for  the optical arc-like feature
3-593.0 is with an Scd galaxy template at  $z$ = 1.73 to 1.76 (Fern\'andez-Soto
et al.  1999; Rowan-Robinson 1999), the redshift likelihood function does have
a much smaller alias as a  starburst galaxy at $z$ = 2.6 (Fern\'andez-Soto,
private communication) to 3.0 (Rowan-Robinson 1999).   The fit is not as good as
for $z$ = 1.73, in particular for the $J$-magnitude, but it is not implausible.
This is in fact the case discussed by Hughes et al. (1998; see their  Fig.~5).
The spectral energy distribution would still match that of VII~Zw~31, as in {\bf
Fig.~6},  scaled up by a factor of 3.6 (to correct for the greater luminosity
distance), except in the optical range, where the arc source 3-593.0 would then
be much more blue than VII~Zw~31.  A better fit for both HDF 850.1 and  3-593.0
would be obtained with the other two sources measured by Trentham et al. (1999),
namely, IRAS 12112+0305 and IRAS F22491$-$1808. Because of the greater luminosity
distance at $z=3$,  the  8\,$\mu$m to 1000\,$\mu$m luminosity would have to be
$L_{\rm IR}$ =  $8.2\times 10^{12}$\,\Lsun , which would place HDF 850.1 among
the most  powerful of the ultraluminous galaxies, like Mrk~231   (see, e.g., the
sample of Sanders et al. 1988).

In summary, the arguments in favor of the interpretation as an 
unlensed source at $z\sim 1.7$  are the fairly good match in both size and  
spectral energy distribution with VII~Zw~31,  
the better fit of the optical template with 
a photometric redshift of 1.7 for 3-593.0, the absence of any strong lines
in the optical spectrum, which is better understood if $z\sim 1.7$, 
the apparent cluster of objects in the range $z =$ 1.6 to 1.8 in this region,
the blue color of the arc, and the fact that the total luminosity 
of $2\times 10^{12}$\,\Lsun\ is not exceptional in the class of 
ultraluminous galaxies, consistent with the absence of any indication 
of an embedded optical quasar or Seyfert~1 nucleus  (see, e.g., Kim \& Sanders
1998; Solomon et al. 1997).

The arguments in favor of an interpretation as an unlensed source at 
$z\sim  3$ have been given by Hughes et al. (1998) and Carilli \& Yun (1999).
To these arguments we may add the poorer-fitting 
alias for a photometric redshift of 2.6 or 3.0.   The consequence is that
the IR luminosity would place the source  among
the most  powerful of the ultraluminous galaxies, those containing 
a quasar or a Seyfert~1 nucleus.  These objects are fairly rare
in the existing samples of ultraluminous galaxies (see the luminosity histogram
of Kim \& Sanders 1998; their Fig.~3), and usually have a small, bright,
symmetric nucleus resembling a reddened QSO (Surace et al. 1998).
There is no evidence for such a bright QSO-like 
nucleus in the optical image of
3-593.0, which instead looks like a star-forming region extended over 
several kpc.

\subsection{The dust source HDF850.1 as a lensed source}
Another possibility is that HDF 850.1 is a lensed source.
The relative abundance of lensed sources in the submm is expected to be large,
because the submm flux density-redshift relations of distant, 
dusty, star-forming galaxies are flat, leading to a larger
fraction of brighter images in the submm relative to the optical (Blain 1997).
The fraction of gravitationally lensed
sources at the 10\,mJy level at 850\,$\mu$m is $> 0.01$ (Blain 1998, his
Fig.~4;   Blain et al. 1999), orders of magnitude greater than for the
corresponding optical sources.    There is therefore a good chance that we are
detecting a lensed source, especially if the line of sight is close to an
elliptical galaxy.

 Hogg et al. (1996) originally proposed that the  arc-like feature 3-593.0
was a source  gravitationally lensed by the elliptical  galaxy  at
$z\approx 1$ because of the  1.8$''$ separation of the arc from the elliptical
3-586, the mass  of the elliptical as deduced from its luminosity, the
appearance of the arc, concave toward the elliptical,   and the fact that other
gravitational lenses tend to be ellipticals.   
It is possible that the dust source is part of the same object
as the optical  arc-like feature. and lies near a
high-magnification point  of the lens. 

The non-detection of optical lines in the arc
is consistent with  $1.2 < z < 2.4$ (S. Zepf, private communication
and Zepf et al. 1997). 
The low end of this range is close to the redshift of
the elliptical, giving an unfavorable geometry for lensing.  The high end of the
range, however, yields nearly the same geometry as  for IRAS F10214+4724, where
the elliptical is at $z=0.9$ and the source is at $z=2.3$. 
If  the optical arc-like feature is at its photometric redshift ($z\sim 1.7$),
then  the arc source and the elliptical are in a favorable configuration for
lensing.   
The approximate radius of the circle of images may be roughly estimated  
 from the standard formula for the 
Einstein radius for an isothermal, spherical lens 
\begin{equation}
\theta_E = \bigg( 
{ 
{\sigma_v} 
\over 
{ 186\,{\rm km\,s}^{-1}  }
} 
\bigg)^2 {D_A(LS) \over D_A(S)}
\end{equation}
(e.g., Peacock 1999, his eq. 4.14), where $\sigma_v$ is the line of sight
velocity dispersion of the lens  in \kms, and $D_A(LS)$ and $D_A(S)$ are the
angular size distances from the  lens to the source and from the observer to the
source respectively.

\begin{table*}
\caption{Comparison of the HDF 850.1/3-586.0/3-593.0 system  
with other lensed sources}
\begin{tabular}{l ccccc cccc}
\hline
&\multicolumn{5}{c}{{--------------- Elliptical galaxy ---------------}} 
&\multicolumn{4}{c}{{--------------- Lensed feature ---------------}}\\			     			     		
&Lens		&$m_{814}$	&$m_{814}$	&$m_{H+K}$	&$\sigma_v$
&Source		&crit.curve	&$m_{814}$  	&opt./UV
\\
Source	
&redshift	&$r<0.2''$	&total		&total		&
&redshift	&radii				&total	  	&magnifi-			
\\
&$z_{\rm lens}$	&(mag)		&(mag)		 &(mag)		&(\kms )	
&$z_s$		&(arcsec)			&(mag)		&cation
\\
\hline
\\
\multicolumn{4}{l}{{\bf HDF 850.1 system:}}
\\
HDF 3-586.0/3-593.0	
&1.0		&---		&23.3		&19.5		&$< 340$		
&$\sim 1.7 ?$	&1.61				&26.1 		&---
\\
\\
\multicolumn{4}{l}{{\bf Other lenses:}}
\\
IRAS F10214+4724	
&0.9		&22.8		&20.3		&18.5		&270 
&2.286		&1.2$\times 0.8$		&20.44	&$\sim 100$
\\
HST 14176+5226	
&0.81		&---		&19.7	&---	&260
&3.4		&$2\times 1.5$			&25.8		&---
\\
Cloverleaf		
&$\sim 1$ ?	&$>24$		&$>22.5$	&$\sim 20.5$	&$<350$
&2.558		&0.7$\times 0.5$		&17.5	&18 to 30
\\
\\
 \hline
\multicolumn{10}{l}{HDF 3-586.0/3-593.0 data from Cowie (1999); 
lens model from Hogg et al. (1996).}\\
\multicolumn{10}{l}{IRAS F10214+4724 data and lens model from Eisenhardt et al. 
(1996); IR data from Evans et al. (1999).}\\
\multicolumn{10}{l}{HST 14176+5226 data from Ratnatunga et al. (1995);
Crampton et al. (1996); lens model from Hjorth \& Kneib (1999).}\\
\multicolumn{10}{l}{Cloverleaf  
data and lens model from Kneib et al. (1998a; 1998b), Turnshek et al. (1997), 
and Chae \& Turnshek (1999).}
\end{tabular}
\end{table*}

In the lower-mass, isothermal, elliptical-potential  
model of Hogg et al. (1996), where only
3-593.0 and  3.577.0 were a candidate lensed pair,   
the optical arc  was close to
the critical radius of 1.6$''$, from which they derived  a velocity dispersion
of $\geq 340$\,\kms \ if the elliptical galaxy is at $z=1$.  If the dust source
is at $z\approx 1.7$, and is a lensed image 0.8$''$ from the elliptical,  then
the very rough estimate from eq.(4) also yields a  velocity dispersion of $\sim
330$\,\kms .   Hogg et al. (1996) tried to explain not only the
apparent  arc 3-593.0, concave toward the elliptical, but also the  object
3-577.0 as its counterimage, elongated toward the elliptical. 
The
observations of Zepf et al. (1997) made this candidate counterimage  unlikely,
however, because of the tentative detection of a Ly$\alpha$ 
line at $z=3.36$ in 3-577.0, 
and the different  $B_{450} -V_{606}$
colors in the arc and the ``counterimage" 
(the arc is bluer than its candidate ``counterimage";    
for an alternative interpretation, see Dickinson 1998).

How could the HDF 850.1 dust source and the optical arc-like feature be  at
different positions?  There are several ways this could happen.  In the source
that is lensed, the dust emission may not be in  the same place as the
UV/optical emission.   The dust source and the optical arc could be different
parts of the same galaxy,   or a merger of two galaxies, with their light paths
deflected  to different spots by  the elliptical. They would intersect the
elliptical's cusp and fold caustics differently, and have different
magnifications, and different image shapes  on the sky.  If the source is at
$z\approx 1.7$ and lensed,  its luminosity would still be that of a starburst
galaxy.    These objects are often  highly distorted, with double nuclei, or
multiple knots of emission ---  good candidates   to yield non-coincident
optical and mm/submm images if gravitationally lensed. 

 Non-coincident images in different wavebands can be 
expected quite naturally, as shown by the comparison with other lensed 
sources in the next section.

\subsection{Comparison with other lensed systems} 
 More than half of the high-redshift objects in which dust  and/or CO
has been detected at 1.3\,mm  are known to be gravitationally lensed, with
magnification factors $\approx10$ for the mm dust emission in some of the
sources   (e.g. IRAS F10214+4724, the Cloverleaf quasar, APM 08279+5255, MG
0414+0534,  and very likely BR 1202$-$07). Other submm-selected sources are also
lensed, with lower amplification factors (e.g.,  SMM J02399-0136 and
SMM J14011+0252: Ivison et al. 1998;  Frayer et al. 1998; 1999).

{\bf Table~2} compares the lower-mass model for the elliptical 3-586.0 from
Hogg et al.\ (1996) with similar models for 
IRAS F10214+4724, HST 14176+5226,
and the Cloverleaf quasar, 
in which the lensing  elliptical galaxies are at comparable redshifts.
An important point in these examples is that  a gravitational lens can 
image  different parts of a source to different spots on the sky in 
different wavebands. 

{\it In IRAS F10214+4724,}  the combination of an elliptical at
$z=0.9$ and a source at $z=2.3$ gives strong magnification with a 1.18$''$
asymptotic critical radius  for the UV  light redshifted to the optical.    The
optical arc in 10214+4724 contains two peaks.  In the model  of Eisenhardt et
al. (1996),  the brighter east peak is interpreted as the blending of two images
merging on the critical curve, while the west peak is interpreted as the third
image.    A possible analogy in the HDF 850.1 system might be the brighter
optical source  3-593.0 as the two-image blend, and the weaker peak 3-593.2 as
the third image. The 10214+4724 counterimage is 100 times fainter than the main
arc,  and has not been detected in the cm-radio or mm-dust emission.
The models by Eisenhardt et al. (1996) for 10214+4724 
predict  a 0.4$''$ radial displacement in the $K$-band arc, 
due to the larger source region at longer wavelengths.  They note 
that an 0.5$''$ infrared source would be imaged into an elliptical 
ring with a position angle perpendicular to that of the optical arc, 
with the center of the ring offset from the elliptical.  
In fact, the HST NICMOS
observations of 10214+4724
by Evans et al. (1999) show that the peak position of the arc
changes even between wavelengths of 1.1 and 2.1\,$\mu$m.

{\it In HST 14176+5226}, the lens redshift, total magnitude of the 
elliptical galaxy in the HST F814W band, and calculated 
line-of-sight velocity dispersion of the elliptical are very similar 
to those of the elliptical that lenses IRAS F10214+4724.
In the detailed model by Hjorth \& Kneib (1999) for the lensing of  
HST 14176+5226, the magnified images in 
the Einstein cross configuration actually lie outside the critical line,
which is an elongated ellipse.  

{\it In the Cloverleaf quasar}, the lensing galaxy has recently been detected
in NICMOS images (Kneib et al. 1998b), and may belong to a galaxy
cluster or group at $z \approx 0.9$.  The lensing effect may be due to a 
combination of the elliptical and the galaxy group.  The different
source size in different spectral regions leads to quite different images.
In the Cloverleaf, the four optical images are circular spots, 
while the 
four millimeter sources are arcs (Kneib et al. 1998a). 

\subsection
{Possible identification of the dust source with the  tentative VLA source
3651+1226 and  implications from the radio-FIR relation} 

From the positions in {\bf Table~1} and {\bf Fig.~4},  the 1.3\,mm dust source
appears to coincide with the weak VLA radio source 3651+1226, listed in the
Supplementary Catalogue of Richards et al. (1998; their Table~5).   At a
wavelength  of 3.5\,cm, this 4.5$\sigma$ source has  a (revised) sky flux
density of $7.5\pm 2.2$\,$\mu$Jy (E.\ Richards, private communication).
From the map by Richards (1999) one may set a 3$\sigma$  upper limit of
23\,$\mu$Jy at 20\,cm. The dust emission extrapolated from the 2\,mJy measured
at 1.3\,mm, would yield only 0.02\,$\mu$Jy at 3.5\,cm,  so the cm-radio source
must be mainly synchrotron emission, not dust.   Using this VLA source in
relation  with  the cm-radio-to-far-IR correlation in low-redshift star-forming
galaxies,   Carilli \& Yun (1999) concluded that HDF 850.1 could 
be at a redshift
$z\geq 3$,  because the 350/1.4\,GHz index $\alpha^{350}_{1.4}$ is $> 1$.    

With regard to this submm-to-radio ratio, it is instructive    to compare the
observed flux density of HDF 850.1 with that expected from the ultraluminous IR
galaxy Arp~220, if the latter were at high redshift.  The result is shown in
{\bf Table~3}, which  indicates that Arp~220 could not be detected much beyond
$z \approx 1.7$  at 20\,cm, even with deep integrations with the VLA, a fact
stressed by Richards (1999).  An unlensed Arp~220 would not be detectable with
the IRAM interferometer at 1.3\,mm to the detection limits reported in this
paper, in the range  $1.7<z<4$.   It would also be 3.5 to 6 times weaker at
850\,$\mu$m  than the flux observed by Hughes et al. (1998) for HDF 850.1.  Nor
would it  have been detected by the ISOCAM instrument at 15\,$\mu$m, in the
same range of redshift.

\begin{table}
\caption{Expected unlensed flux densities of Arp~220 at high-$z$}
\begin{tabular}{c cccc}
\hline
&\multicolumn{4}{l}{Arp 220 expected flux at wavelength:}\\
		&VLA		&IRAM		&SCUBA		&ISOCAM\\
Redshift 	&20\,cm		&1.3\,mm	&850\,$\mu$m	&15\,$\mu$m \\
$z$		&($\mu$Jy)	&(mJy)		&(mJy)		&($\mu$Jy)\\ 			     			     
\hline
0.01818		&$3\times 10^5$	&180		&800	&$8\times 10^5$\\
1.7		&37		&0.3		&1.2		&18\\
3.0		&14		&0.5		&2.0		&1.7\\
4.0		&8		&0.8		&2.1		&0.8\\
\hline
{\bf HDF 850.1}	\\
{\bf observed}	&{\bf $<$ 23}	&{\bf 2.2}	&{\bf 7.0}	&{\bf $<$ 23}\\
\hline   
\multicolumn{5}{l}{
Values are scaled to Arp 220's observed spectrum:}\\ 
\multicolumn{5}{l}{
{\it radio}: Sopp \& Alexander (1991),}\\ 
\multicolumn{5}{l}{
{\it millimeter}: Downes \& Solomon (1998),}\\
\multicolumn{5}{l}{
{\it far-IR}: Rigopoulou et al. (1996); Fischer et al. (1998);}\\ 
\multicolumn{5}{l}{Klaas et al. (1997),
{\it near-IR}: Soifer et al. (1999).}\\
\end{tabular}
\end{table}

As noted by Carilli \& Yun (1999), however, the 
submm-to-radio index has a large scatter, so it should be regarded 
more as a redshift indicator rather than a redshift estimator in the sense 
of the optical photometric redshifts.  In particular, for the high-$z$ dust
sources detected to date,  there is not only  a large scatter in the
350/1.4\,GHz index, but also possible contamination of the radio flux  by
emission from an AGN, and modification of the ratio due to  differential
magnification effects of gravitational lensing. For example, IRAS F10214+4724,
at $z = 2.3$,   has  an index $\alpha^{350}_{1.4}$ = 0.7, and APM 08279+5255, at
higher redshift ($z=3.9$) has  nearly the same index,  $\alpha^{350}_{1.4}$ =
0.8.  Both of these indices would be compatible with $z<2$, given the scatter in
the index data.

\section{Conclusions}
1) The IRAM interferometer has detected the source HDF 850.1, the strongest
source in the submm survey of the Hubble Deep Field 
by Hughes et al. (1998).  We measure a flux of $2.2\pm 0.3$\,mJy, 
 in agreement with  the flux density 
measured at 1.35\,mm at the James Clerk Maxwell Telescope.
 The 1.3\,mm flux, together with our non-detections at 2.8 and 3.4\,mm, and
the SCUBA flux at 850\,$\mu$m, are all consistent with the interpretation by 
Hughes et al. (1998) that
the radiation is optically thin emission by dust.

2) The improved  position from the interferometer shows
the dust source cannot be the  optical source 
3-577.0, as suggested by Hughes et al. (1998).  
The measured position also rules out 
 the $z=0.299$ galaxy 3-659.1 and the 
low surface brightness galaxy 3-633.1 at $z=1.72$  suggested by 
Richards (1999) as an indentification for HDF 850.1. The dust source may 
coincide with the tentative VLA source 3651+1226.

3) The dust source  may be associated with the optical arc-like feature
3-593.0/3-593.2, located  1.8$''$ to the southwest of the $z\approx 1$
elliptical galaxy 3-586.0.
The arguments in favor of this interpretation are
the $0.8'' \pm  0.7''$   separation of the dust source and 3-593 on the sky
and 
the temperature-mass argument developed in the text, which disfavors 
an association of the dust source with the elliptical. 
If HDF 850.1 is at $z\approx 1.7$, the photometric redshift
of the optical source, then it could be an unlensed 
galaxy with a spectral energy distribution and  
linear dimensions that are similar to those of the $z=0.054$
ultraluminous starburst galaxy VII~Zw~31.  Hence, if at $z\approx 1.7$ and 
unlensed, HDF 850.1
must be an ultraluminous IR galaxy, with an IR luminosity of   
 $L_{\rm IR}$ = $2\times 10^{12}$\,\Lsun .

4) Because of the close proximity on the sky of the elliptical 3-586.0, the
dust source    
 HDF 850.1 may be gravitationally lensed.
The arguments in favor of this interpretation are
 the arclike morphology of the optical source 
that suggests a gravitational mirage;
the similarity of the properties of the elliptical galaxy to the ones 
responsible
for the lensing in IRAS F10214+4724, HST 14176+5226, and the Cloverleaf; 
and the fact that many of the other high-redshift objects detected to date
at mm/sub-mm wavelengths in dust and CO are magnified by 
gravitational lensing.   The arguments against this interpretation are 
that the arc morphology is not conclusive, and that the candidate 
counterimage, 3-577.0, has a different $B-V$ color, and apparently a 
different redshift.  

 The discussion in this paper illustrates the absolute necessity of
sensitive millimeter/submillimeter interferometers for identifying 
sources of dust emission with cosmologically distant optical galaxies.
The next step in this direction is the construction of a mm/submm 
interferometer with more than an order of magnitude greater sensitivity than 
presently available.

\begin{acknowledgements} 
We thank the operators on Plateau de Bure for their help in observing,
M.~Bremer (IRAM) for some of the image processing, and   
H.~ Aussel (CEA-Saclay),
M.~Dickinson (STSci),
A.~Fern\'andez-Soto (Univ.\ New South Wales), 
J.-P.~Kneib (Obs.\ Midi-Pyr\'en\'ees), 
E.~Richards (Univ.\ Virginia),  P.A.~Shaver (ESO), and 
S.~Zepf  (Yale Univ.) for
very useful comments.  We also thank the referee, A.~Omont, for 
his helpful suggestions on improving the paper. 
\end{acknowledgements}

\noindent
{\bf Appendix: The cm-radio source 3651+1221}

\noindent
Richards et al. (1998) tentatively associated the  VLA source 3651+1221  with
the $z=0.299$ galaxy 3-659.1   (cf. {\bf Fig.~4}). This galaxy may also
be the 15\,$\mu$m source HDF\_PM3\_29 detected with ISOCAM (Aussel et al. 1999).
Richards (1999) showed however that the radio emission is confined to  the
faint emission north of the spiral, about $1''$ south of the optical   object
3-633.1, which has a photometric redshift of 1.72 (Fern\'andez-Soto et al. 1999).
Since then, this optical object has been detected in $J$ and $H$ bands, having
colors consistent with a high-$z$ galaxy (Dickinson 1998).
Richards (1999) noted that    if the radio source is at $z\sim 1.7$, the implied
radio luminosity is  substantially higher than that of the most extreme local
starbursts like  Arp~220, which suggests the radio source may be an AGN.

There is a 3-sigma contour near the spiral galaxy on the IRAM 1.3\,mm map ({\bf
Fig.~4}), which we list in {\bf Table~1} as a  questionable source.   The
1.3\,mm position is 1$''$ south of the VLA source. The apparent 1.3\,mm flux
density is 0.75\,mJy, but this would have to be corrected for the primary beam
attenuation to $\sim 1.2$\,mJy if the source were real.    In this case,
the emission could not be the extrapolation of the synchrotron spectrum from
centimeter  wavelengths, because the flux density at 3.4\,cm is only 16\,$\mu$Jy
(Richards  et al. 1998).  Could it be dust emission from the spiral?  The
problem with dust  emission at low redshifts ($z\sim 0.3$) is that most of the
energy is at shorter wavelengths,  not redshifted into the mm /submm bands.
The dust in the central parts of the Milky Way, or in  
NGC 891, corresponding to a molecular gas mass of $1.5\times 10^9$\,\Msun ,
would yield a 1.3\,mm flux of only 68\,$\mu$Jy at $z= 0.299$,
or $\sim 20$ times lower than the apparent flux  on the 1.3\,mm map.
The apparent 3$\sigma$ signal  would thus have to come from  
an extremely gas-rich galaxy, which is not in any way 
 indicated by the HST optical image of this relatively nearby object.
Since there is also a  3$\sigma$ negative peak at a similar radius from
the center of the primary beam ({\bf Fig.~1}), the apparent 3$\sigma$ 
positive signal may not be real.


\begin{thebibliography}{}
\bibitem[]{}
Aussel, H., Cesarsky, C.J., Elbaz, D., \& Starck, J.L.  1999, A\&A, 342,
313
\bibitem[]{}
Barger, A.J., Cowie, L.L., \& Sanders, D.B. 1999a, ApJ, 518, (June 10) in press
\bibitem[]{}
Barger, A.J., Cowie, L.L., Smail, I., Ivison, R.J., Blain, A.W., \& Kneib, 
J.-P. 1999b, AJ, (June 1999) in press (astro-ph/9903142)
\bibitem[]{}
Blain, A.W. 1997, MNRAS, 290, 553
\bibitem[]{}
Blain, A.W. 1998, MNRAS, 295, 92
\bibitem[]{}
Blain, A.W., Kneib, J.-P., Ivison, R.J., \& Smail, I. 1999, ApJ, 512, L87
\bibitem[]{}
Carilli, C.L., \& Yun, M.S. 1999, ApJ, 513, L13
\bibitem[]{}
Chae, K.-H., \& Turnshek, D.A. 1999, ApJ, 514, 587 
\bibitem[]{}
Cimatti, A., Andreani, P., R\"ottgering, H., \& Tilanus, R. 1998, 
Nature, 392, 895
\bibitem[]{}
Clark, B.G. 1980, A\&A, 89, 377
\bibitem[]{}
Condon, J.J., Helou, G., Sanders, D.B., \& Soifer, B.T. 1996, ApJS, 103, 81
\bibitem[]{}
Cowie, L. 1999, http://www.ifa.hawaii.edu/$\sim$cowie/tts /hdf42.html
\bibitem[]{}
Crampton, D., Le F\`evre, O., Hammer, F., \& Lilly, S.J. 1996, A\&A, 307, L53
\bibitem[]{}
Dey, A., Graham, J.R., Ivison, R.J., Smail, I., Wright, G.S., \& Liu, M.C.
1999, ApJ, in press (astro-ph/9902044)
\bibitem[]{}
Dickinson, M. 1998, in The Hubble Deep Field, ed. M.\ Livio, S.M.\ Fall, 
\& P.\ Madau, Cambridge: Cambridge Univ.\ Press, 219
\bibitem[]{}
Djorgovski, S., de Carvalho, R.R., \& Thompson, D.J. 1990, AJ, 99, 1414
\bibitem[]{}
Downes, D., Radford, S.J.E., Greve, A., Thum, C., Solomon, P.M., \& Wink, J.E.
1992, ApJ, 398, L25
\bibitem[]{}
Downes, D., Solomon, P.M., \& Radford, S.J.E. 1993, ApJ, 414, L13
\bibitem[]{}
Downes, D., \& Solomon, P.M. 1998, ApJ, 507, 615
\bibitem[]{}
Eales, S., Lilly, S., Gear, W., Dunne, L., Bond, J.R., Hammer, F., 
Le~F\`evre, O., \& Crampton, D.  1999, ApJ, 515, 518
\bibitem[]{}
Eisenhardt, P.R., Armus, L., Hogg, D.W., Soifer, B.T., Neugebauer, G.,
\& Werner, M.W. 1996, ApJ, 461, 72
\bibitem[]{}
Evans, A.S., Scoville, N.Z., Dinshaw, N., Armus, L., Soifer, B.T., 
Neugebauer, G., \& Rieke, M. 1999, ApJ, in press (astro-ph/9812196)
\bibitem[]{}
Fasano, G., Christiani, S., Arnouts, S., \& Filippi, M.
1998, AJ, 115, 1400
\bibitem[]{}
Fern\'andez-Soto, A., Lanzetta, K.M., \& Yahil, A. 1999, ApJ, 513, 34
(http://bat.phys.unsw.edu.au/$\sim$fsoto/hdfcat.html)
\bibitem[]{}
Fischer, J., Satyapal, S., Luhman, M.L., Melnick, G., Cox, P., Cernicharo, J.,
Stacey, G.J., Smith, H.A., Lord, S.D., \& Greenhouse, M.A.
1998, in ISO to the Peaks, ed., M.\ Kessler \& M.\ Perry, 
Noordwijk: ESTEC, in press
\bibitem[]{}
Frayer, D.T., Ivison, R.J., Scoville, N.Z., Yun, M.,  Evans, A.S., Smail, I.,
Blain, A.W., \& Kneib, J.-P. 1998, ApJ, 506, L7
\bibitem[]{}
Frayer, D.T., Ivison, R.J., Scoville, N.Z., Evans, A.S., Yun, M.S., Smail, I.,
Barger, A.J., Blain, A.W., \& Kneib, J.-P. 1999, ApJ, 514, L13
\bibitem[]{}
Gu\'elin, M., Zylka, R., Mezger, P.G., Haslam, C.G.T., Kreysa, E., Lemke, R.,
\& Sievers, A.W. 1993, A\&A 279, L37
\bibitem[]{}
Gwyn, S.D.J., \& Hartwick, F.D.A. 1996, ApJ, 468, L77
(http:/uvastro.phys.uvic.ca/grads/gwyn)
\bibitem[]{}
Hjorth, J., \& Kneib, J.-P. 1999, ApJ, in press
\bibitem[]{}
Hogg, D.W., Blandford, R., Kundi\'c, T., Fassnacht, C.D., \&
Malhotra, S. 1996, ApJ, 467, L73
\bibitem[]{}
Hughes, D.H., Dunlop, J.S., \& Rawlings, S. 1997, MNRAS, 289, 766
\bibitem[]{}
Hughes, D., et al. 1998, Nature, 394, 241
\bibitem[]{}
Ivison, R.J., Smail, I., Le~Borgne, J.-F., Blain, A.W., Kneib, J.-P.,
\& B\'ezecourt, J., Kerr, T.H., \& Davies, J.K. 1998, MNRAS, 298, 583
\bibitem[]{}
Kim, D.-C., \& Sanders, D.B. 1998, ApJS, 119, 41
\bibitem[]{}
Klaas, U., Haas, M., Heinrichsen, I., \& Schulz, B. 1997, A\&A, 325, L21
\bibitem[]{}
Knapp, G.R. 1999, in Star Formation in Early-Type Galaxies, ed. P.\ Carral
\& J.\ Cepa, San Francisco: Astron. Soc. Pacific, in press
\bibitem[]{}
Kneib, J.-P., Alloin, D., Mellier, Y., Guilloteau, S., Barvainis, R., 
\& Antonucci, R. 1998a, A\&A, 329, 827
\bibitem[]{}
Kneib, J.-P., Alloin, D., \& Pell\'o, R.  1998b, A\&A, 339, L65
\bibitem[]{}
Kr\"ugel, E., \& Siebenmorgen, R. 1994, A\&A, 288, 929 
\bibitem[]{}
Lanzetta, K.M., Yahil, A., \& Fern\'andez-Soto, A. 1996, Nature, 381, 759
\bibitem[]{}
Lilly, S.J., Eales, S.A., Gear, W.K.P., Hammer, F., Le~F\`evre, O., 
Crampton, D., Bond, J.R., \& Dunne, L. 1999, ApJ, in press (astro-ph/9901047)
\bibitem[]{}
Mann, R.G., et al. 1997, MNRAS, 289, 482
\bibitem[]{}
Mazzarella, J.M., et al. 1999, in preparation
\bibitem[]{}
Mobasher, B., Rowan-Robinson, M., Georgakakis, A., \& Eaton, N. 1996, 
MNRAS, 282, L7
\bibitem[]{}
Patnaik, A.R., Browne, I.W.A., Wilkinson, P.N., \& Wrobel, J.M.
1992, MNRAS, 254, 655
\bibitem[]{}
Peacock, J.A. 1999, Cosmological Physics, Cambridge: Cambridge Univ.\ Press, 106
\bibitem[]{}
Ratnatunga, K.U., Ostrander, E.J., Griffiths, R.E., \& Im, M. 1995, 
ApJ, 453, L5
\bibitem[]{}
Richards, E.A. 1999, ApJ, 513, L9 
 
(http://www.cv.nrao.edu/$\sim$jkempner/vla-hdf)
\bibitem[]{}
Richards, E.A., Kellermann, K.I., Fomalont, E.B., Windhorst, R.A., 
\& Partridge, R.B. 1998, AJ, 116, 1039
\bibitem[]{}
Rigopoulou, D., Lawrence, A., \& Rowan-Robinson, M. 1996, MNRAS, 278, 1049
\bibitem[]{}
Rivolo, A.R., \& Solomon, P.M.  1988, in Molecular Clouds in the 
Milky Way and External Galaxies, ed. R.L.\ Dickman, R.L.\ Snell, \& J.S. Young,
Heidelberg: Springer, 42
\bibitem[]{}
Rowan-Robinson, M. 1999, in preparation
\bibitem[]{}
Rowan-Robinson, M., et al. 1997, MNRAS, 289, 490
\bibitem[]{}
Sanders, D.B., Soifer, B.T., Elias, J.H., Madore, B.F., Matthews, K., 
Neugebauer, G., \& Scoville, N.Z. 1988, ApJ, 325, 74
\bibitem[]{}
Sawicki, M.J., Lin, H., \& Yee, H.K.C.  1997, AJ, 113, 1; 
(http://www.astro.utoronto.ca/$\sim$sawicki/)
\bibitem[]{}
Soifer, B.T., Neugebauer, G., Matthews, K., Becklin, E.E., Ressler, M.,
Werner, M.W., Weinberger, A.J., \& Egami, E.
1999, ApJ, 513, 207 
\bibitem[]{}
Solomon, P.M., Downes, D., Radford, S.J.E., \& Barrett, J.W. 1997, ApJ, 478, 144
\bibitem[]{}
Sopp, H.M., \& Alexander, P. 1991, MNRAS, 251, 112
\bibitem[]{}
Surace, J.A., Sanders, D.B., Vacca, W.D., Veilleux, S., \& Mazzarella, J.M.
1998, ApJ, 492, 116
\bibitem[]{}
Trentham, N., Kormendy, J., \& Sanders, D.B. 1999, AJ, in press
(astro-ph/9901382)
\bibitem[]{}
Turnshek, D.A., Lupie, O.L., Rao, S.M., Espey, B.R., \& Sirola, C.J.
1997, ApJ, 485, 100
\bibitem[]{}
Wang, Y., Bahcall, N., \& Turner, E.L. 1998, AJ, 116, 2081; 
ftp.astro.princeton.edu, cd elt/:HDF
\bibitem[]{}
Wiklind, T., \& Henkel, C. 1995, A\&A, 297, L71
\bibitem[]{}
Williams, R.E., et al. 1996, AJ, 112, 1335
\bibitem[]{}
Wilner, D.J., \& Wright, M.C.H. 1997, ApJ, 488, L67
\bibitem[]{}
Wink, J.E., Duvert, G., Guilloteau, S., G\"usten, R., Walmsley, C.M., 
\& Wilson, T.L. 1994, A\&A, 281, 505
\bibitem[]{}
Wyrowski, F., Hofner, P., Schilke, P., Walmsley, C.M., Wilner, D.J., 
\& Wink, J.E. 1997, A\&A, 320, L17
\bibitem[]{}
Wyrowski, F., Schilke, P., Walmsley, C.M., \& Menten, K.M.
1999, ApJ, 514, L43  
\bibitem[]{}
Zepf, S.E., Moustakas, L.A., \& Davis, M. 1997, ApJ, 474, L1
\end{thebibliography}
\end{document}